\documentclass[twocolumn]{aastex63}

\accepted{for publication in ApJL}

\newcommand{\OII}{[\ion{O}{2}]}

\newcommand{\Msun}{\ensuremath{M_{\odot}}}
\newcommand{\Mstar}{\ensuremath{M_{\ast}}}
\newcommand{\Mgas}{\ensuremath{M_\mathrm{gas}}}
\newcommand{\HST}{\emph{HST}}
\newcommand{\Spitzer}{\emph{Spitzer}}

\shorttitle{Molecular Gas in Quenching Galaxies at $z\sim1$}
\shortauthors{Belli et al.}

\begin{document}

\title{The Diverse Molecular Gas Content of Massive Galaxies Undergoing Quenching at $z\sim1$}

\author{Sirio Belli}
\affiliation{Center for Astrophysics $|$ Harvard \& Smithsonian, Cambridge, MA, USA}

\author{Alessandra Contursi}
\affiliation{Max-Planck-Institut für Extraterrestrische Physik, Garching, Germany}
\affiliation{Institut de Radioastronomie Millimétrique, Saint Martin d'Hères, France}

\author{Reinhard Genzel}
\author{Linda J. Tacconi}
\author{Natascha M. F\"orster-Schreiber}
\author{Dieter Lutz}
\affiliation{Max-Planck-Institut für Extraterrestrische Physik, Garching, Germany}

\author{Françoise Combes}
\affiliation{Observatoire de Paris, LERMA, College de France, CNRS, PSL, Sorbonne University,  Paris, France}

\author{Roberto Neri}
\affiliation{Institut de Radioastronomie Millimétrique, Saint Martin d'Hères, France}

\author{Santiago García-Burillo}
\affiliation{Observatorio Astronómico Nacional (OAN)-Observatorio de Madrid, Madrid, Spain}

\author{Karl F. Schuster}
\affiliation{Institut de Radioastronomie Millimétrique, Saint Martin d'Hères, France}

\author{Rodrigo Herrera-Camus}
\affiliation{Universidad de Concepción, Barrio Universitario, Concepción, Chile}

\author{Ken-ichi Tadaki}
\affiliation{National Astronomical Observatory of Japan, Mitaka, Tokyo, Japan}

\author{Rebecca L. Davies}
\affiliation{Max-Planck-Institut für Extraterrestrische Physik, Garching, Germany}
\affiliation{Swinburne University of Technology, Melbourne, Australia}

\author{Richard I. Davies}
\affiliation{Max-Planck-Institut für Extraterrestrische Physik, Garching, Germany}

\author{Benjamin D. Johnson}
\affiliation{Center for Astrophysics $|$ Harvard \& Smithsonian, Cambridge, MA, USA}

\author{Minju M. Lee}
\affiliation{Max-Planck-Institut für Extraterrestrische Physik, Garching, Germany}

\author{Joel Leja}
\affiliation{Center for Astrophysics $|$ Harvard \& Smithsonian, Cambridge, MA, USA}
\affiliation{The Pennsylvania State University, University Park, PA, USA}

\author{Erica J. Nelson}
\affiliation{University of Colorado, Boulder, CO, USA}

\author{Sedona H. Price}
\author{Jinyi Shangguan}
\author{T. Taro Shimizu}
\affiliation{Max-Planck-Institut für Extraterrestrische Physik, Garching, Germany}

\author{Sandro Tacchella}
\affiliation{Center for Astrophysics $|$ Harvard \& Smithsonian, Cambridge, MA, USA}

\author{Hannah \"Ubler}
\affiliation{Max-Planck-Institut für Extraterrestrische Physik, Garching, Germany}

\begin{abstract}
We present a detailed study of the molecular gas content and stellar population properties of three massive galaxies at $1 < z < 1.3$ that are in different stages of quenching. The galaxies were selected to have a quiescent optical/near-infrared spectral energy distribution and a relatively bright emission at 24 \micron, and show remarkably diverse properties.
CO emission from each of the three galaxies is detected in deep NOEMA observations, allowing us to derive molecular gas fractions $\Mgas/\Mstar$ of 13-23\%. We also reconstruct the star formation histories by fitting models to the observed photometry and optical spectroscopy, finding evidence for recent rejuvenation in one object, slow quenching in another, and rapid quenching in the third system. To better constrain the quenching mechanism we explore the depletion times for our sample and other similar samples at $z\sim0.7$ from the literature. We find that the depletion times are highly dependent on the method adopted to measure the star formation rate: using the UV+IR luminosity we obtain depletion times about 6 times shorter than those derived using dust-corrected [OII] emission. When adopting the star formation rates from spectral fitting, which are arguably more robust, we find that recently quenched galaxies and star-forming galaxies have similar depletion times, while older quiescent systems have longer depletion times. These results offer new, important constraints for physical models of galaxy quenching.
\end{abstract}

\keywords{Galaxy quenching --- High-redshift galaxies --- Molecular gas --- CO line emission --- Galaxy stellar content}


\section{Introduction}
\label{sec:intro}

In order to reproduce the observed population of massive quiescent galaxies, models of galaxy formation require the introduction of a \emph{quenching} mechanism that shuts off star formation in massive systems. In cosmological simulations, this is typically achieved via feedback from Active Galactic Nuclei (AGN), but conclusive observational evidence in favor of this scenario is still lacking \citep[][and references therein]{harrison17}. Moreover, the observed diversity of quiescent galaxies in terms of their structure and stellar populations suggests the existence of more than one quenching mechanism \citep[e.g.][]{schawinski14}. Notably, a rapid quenching channel is required to explain the existence of post-starburst galaxies, whose spectral features can only be produced if star formation is shut off in a few hundred Myr or less.
This is substantially shorter than the timescale obtained in some quenching models based on AGN feedback \citep[e.g.,][]{wright19}, and may require gas-rich events such as mergers or violent disk instabilities \citep[e.g.,][]{dekel14}. While post-starburst galaxies make up a small fraction of the total population in the local universe, their incidence appears to increase with redshift, and they likely account for the majority of the quenching population at $z>2$ \citep{wild16, belli19}.

In addition to the quenching timescale, there are other physical properties that can be used to discriminate between competing models. Among the most important ones is the content of molecular gas, which represents the fuel for star formation. In some quenching mechanisms (e.g., gas heating from virial shocks or AGN feedback) the star formation decline is directly caused by a lack of molecular gas; while in others (e.g., gravitational shear by a stellar bulge or turbulence injection by AGN) the cold gas is unable to collapse and form new stars. By observing the molecular gas content in quiescent galaxies at high redshift, close to the epoch of quenching, we can therefore place a strong constraint on the quenching processes. This is, however, observationally challenging due to the small amount of gas present in these systems, and to the additional requirement of deep rest-frame optical data, which are needed to characterize the star formation history and place each galaxy in the correct evolutionary context. Recent studies have finally obtained CO detections for a small number of quiescent galaxies at intermediate redshift \citep{suess17, spilker18}, while most attempts at $z>1$ have led to non-detections \citep{sargent15, bezanson19}.

In this work, we present new observations of molecular gas emission carried out with the Northern Extended Millimeter Array (NOEMA) for three galaxies in different stages of quenching at $1 < z < 1.3$. Combined with deep optical spectra from the W.M. Keck Observatory, these data offer a unique view on the role of gas consumption in galaxy quenching at high redshift.


\begin{figure*}[ht]
\centering
\includegraphics[width=\textwidth]{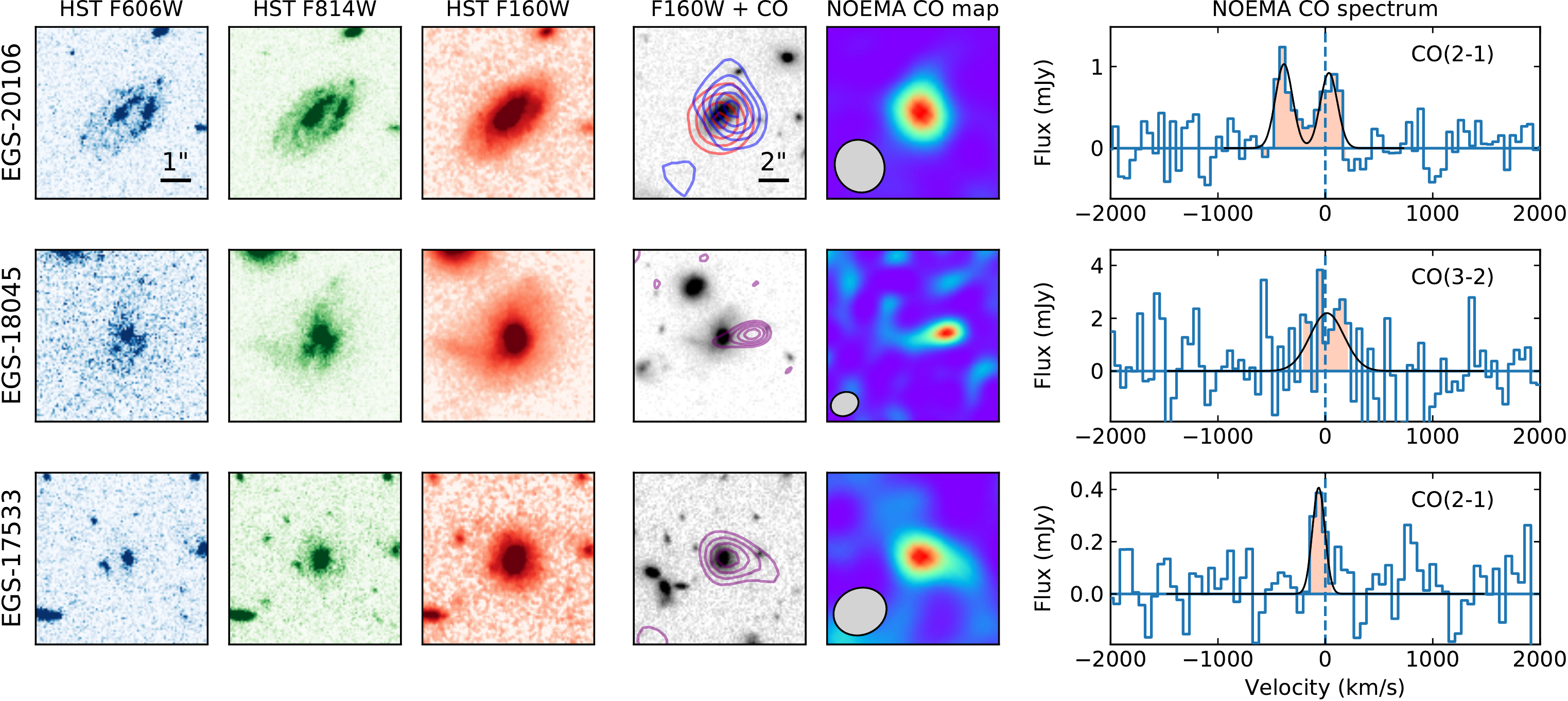}
\caption{\HST\ and NOEMA data for our sample. For each galaxy, the first three panels are $6\arcsec \times 6 \arcsec$ and show the \HST\ images from ACS/F606W, ACS/F814W, and WFC3/F160W. The next two panels are $12\arcsec \times 12 \arcsec$ and show the F160W image with the CO contours in steps of 1$\sigma$ overlaid (starting from the 2-$\sigma$ level and, if the line is spectrally resolved, split into the red and blue sides of the emission); and the CO integrated map, with the beam shown in gray. The last panel shows the NOEMA spectrum of the CO emission line, together with the Gaussian fit (black line). \label{fig:cutouts}}
\end{figure*}

\section{Data}
\label{sec:data}

\subsection{Sample Selection and Ancillary Data}

The targets were selected from a sample of massive quiescent galaxies at $1<z<1.5$ with publicly available optical spectroscopy from deep Keck/LRIS observations \citep{newman10,belli14lris}. In order to ensure a detection of the CO emission line, we decided to target only quiescent galaxies with a relatively strong emission in the infrared. This selection is partly motivated by the results of \citet{spilker18}, who targeted eight quiescent galaxies at $z\sim0.7$ and detected CO emission only in those systems with slightly brighter IR emission, suggesting that these galaxies are not fully quenched yet, as confirmed by an analysis of their rest-frame colors and optical spectra.

We therefore selected galaxies from the Keck sample for which the star formation rate (SFR) obtained from a fit to the broadband spectral energy distribution (SED) is at least a factor of 10 below the main sequence of star formation, but with a total infrared luminosity (inferred from the \Spitzer/MIPS 24 \micron\ photometry) brighter than $10^{11}~L_\odot$.
We also require a location in the EGS field, which has good visibility from the northern hemisphere and a rich set of ancillary data, including \emph{Hubble Space Telescope} (\HST) imaging from the CANDELS survey \citep{grogin11, koekemoer11}; \HST\ grism spectroscopy from the 3D-HST survey \citep{momcheva16}; and multi-band photometry spanning from the UV to the near-IR collected by \citet{skelton14}. 

This selection yields seven galaxies, with stellar masses in the range $10.8 < \log \Mstar/\Msun < 11.3$. We exclude a strong contamination of the sample from AGNs, since none of the targets have \emph{Spitzer} IRAC colors that are near the region populated by AGNs \citep{donley12}. 
Out of these seven galaxies, one has publicly available millimeter data from the PHIBSS2 survey \citep{freundlich19}, and we obtained new observations for two more systems, choosing those with a combination of strong 24 \micron\ emission and high signal-to-noise ratio (due to a longer exposure) in the Keck spectrum.

\subsection{NOEMA Observations}

Using the NOEMA array we targeted the CO(2-1) transition in two galaxies: EGS-20106 was observed in 2019 with an on-source integration time of 11 hours, while EGS-17533 was observed in 2020 with an on-source integration time of 15 hours. The observations were carried out in the 3 mm band, using 10 antennas in the D configuration, which is the most compact. The third object in the sample, EGS-18045, was observed with NOEMA as part of the PHIBSS2 survey in the 2 mm band, targeting the CO(3-2) emission, for 11 hours with 6 antennas in the C and D configurations.

The secondary flux calibrator MWC349, whose flux is regularly measured using planets, was used to derive the absolute flux scale.
The data were calibrated with the \texttt{CLIC} package and mapped with the \texttt{MAPPING} package in the \texttt{GILDAS} software\footnote{\url{http://www.iram.fr/IRAMFR/GILDAS/}}. We used the \texttt{CLARK} cleaning algorithm in a support around the targets, adopting natural weighting, and reconstructed images with beam size of $\sim 2\arcsec - 3\arcsec$. The absolute flux calibration is accurate at the 10\% level.


\section{Analysis}
\label{sec:analysis}

\begin{figure*}[htbp]
\centering
\includegraphics[width=\textwidth]{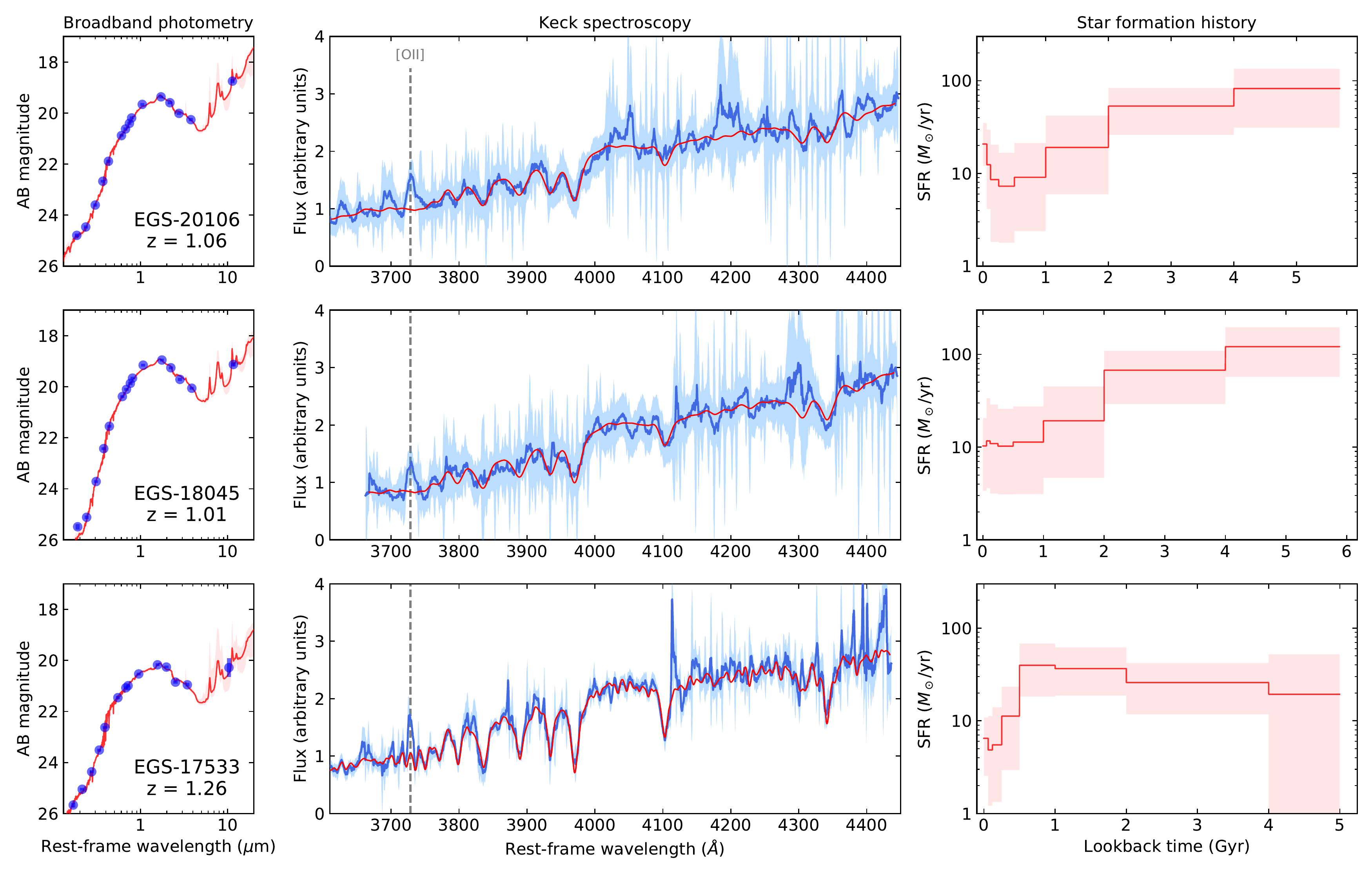}
\caption{Results of the \texttt{Prospector} fits. For each galaxy, the three panels show respectively the observed SED (blue points), the Keck spectrum (blue line, uncertainty in light blue), and the star formation history. In each panel the model is shown in red, with the shaded area showing 90\%\ of the posterior distribution. For illustration purposes, the Keck spectra are inverse-variance smoothed. The location of the \OII\ emission line is marked by a gray dashed line. \label{fig:fit}}
\end{figure*}

\subsection{CO Emission and Molecular Gas Masses}
\label{sec:CO}

Figure~\ref{fig:cutouts} shows the \HST\ cutouts, the map of the CO integrated flux, and the extracted CO spectrum for each galaxy in our sample. We use the spectroscopic redshifts from the Keck spectra, which have an uncertainty of 50 km/s or better, to set the velocity zero-point for the NOEMA spectra. The CO line fluxes are measured by fitting one or two Gaussians to the data, but we obtain consistent results when directly integrating the observed fluxes.

Despite its small size, this sample is characterized by a diversity of CO line properties and optical/near-IR morphology. The CO emission in \textbf{EGS-20106} shows two clear peaks which are resolved both spectrally and spatially. The red peak matches the redshift from the Keck spectrum, while the other peak is blueshifted by about 500 km/s. Interestingly, the spatial location of the blue peak roughly corresponds to that of a bright UV clump, suggesting that the CO line profile is likely not due to regular rotation, but rather to two distinct regions of emission: one in the center of the galaxy and one in what could be a clump or, given the large velocity difference, a gas-rich satellite. Given the small spatial separation, we are unable to carry out a separate analysis of the two components, and we caution that the measured physical properties (such as gas mass and SFR) should be interpreted as the sum of the two components.

\textbf{EGS-18045} has shallower millimeter data compared to the rest of the sample since it was observed with a less powerful array, and the CO line is detected only at the 3-$\sigma$ level; however this detection is highly unlikely to be spurious given that its redshift matches that of the optical spectrum. This galaxy has a clumpy UV morphology and a large tidal tail, likely due to a recent interaction or merger. The CO emission is offset by $2 \arcsec$ from the center of the near-IR continuum, possibly because of the interaction. Alternatively, the CO emission may originate in a companion that is so dust-obscured to be invisible in the \HST\ images, similarly to the system discovered by \citet{schreiber18jekyll} (although we would expect to detect a large offset in velocity as well). In this case the gas mass we measure for EGS-18045 should be considered as an upper limit.

Finally, the CO emission line in \textbf{EGS-17533} is spectrally and spatially unresolved, with a measured velocity dispersion of only $58 \pm 15$ km/s. This is comparable to the velocity dispersion measured from the stellar absorption lines \citep[$88 \pm 18$ km/s,][]{belli14lris}, but significantly lower than the value expected for a pressure-supported system, which is $\sigma_\mathrm{virial} > 130$ km/s (this is a lower limit obtained by neglecting the dark matter contribution). The most likely explanation for this discrepancy is that both the stellar and the molecular gas content are distributed in a nearly face-on rotating disk, which is also consistent with the circular morphology of the galaxy (see \citealt{mowla19} for similar cases). Such alignment suggests that the molecular gas reservoir was formed together with the stellar content, and is not a product of recent accretion. 

To estimate the molecular gas masses from the luminosity of the CO line we adopt the standard conversion factor $\alpha_\mathrm{CO} = 4.4~\Msun$ / (K km s$^{-1}$ pc$^2$) \citep{bolatto13}. Since this factor is calibrated on the 1-0 transition, we also need to assume a brightness temperature ratio $R_{J1}$ describing the CO excitation compared to the thermalized case. Studies of local and high-redshift galaxies suggest a relatively small range of values for low-$J$ transitions \citep[e.g.,][]{carilli13}; for consistency with the PHIBSS survey we adopt $R_{21} = 0.77$ and $R_{31} = 0.50$. The gas masses are in the range $\log \Mgas/\Msun \sim 9.9 - 10.7$, and are listed in Table~\ref{tab:sample} together with other properties of the sample.

\subsection{UV-to-IR Spectral Fit}
 
We explore the stellar population properties of our sample by fitting synthetic templates simultaneously to the continuum-normalized Keck spectroscopy (over the rest-frame wavelengths between 3610~\AA\ and 4450~\AA) and the broadband photometry (which includes UV, optical, near-infrared, and MIPS 24~$\micron$ observations).
We adopt templates from the Flexible Stellar Population Synthesis library \citep[FSPS;][]{conroy09, conroy10} with a \citet{chabrier03} initial mass function, and use the \texttt{Prospector} code \citep{johnson20} together with \texttt{dynesty} \citep{Speagle20} to perform an efficient sampling of the high-dimensional parameter space.
The physical model is based on the \texttt{Prospector-$\alpha$} model \citep{leja17}, and consists of 19 free parameters describing the contribution of stars, dust, and systematic effects to the observed emission. While not all 19 parameters are constrained by the data, the use of a highly flexible model together with physically motivated priors prevents the results from being dominated by our assumptions. The stellar population is described by a set of basic properties (redshift, velocity dispersion, metallicity, mass) and a non-parametric star formation history consisting of seven independent age bins logarithmically spaced; we adopt the continuity prior of \citet{leja19nonparametric} which tends to favor a smooth variation of the SFR from one bin to the next. Absorption from the diffuse dust is described by the 3-parameter model of \citet{kriek13}; the absorbed energy is then re-radiated in the infrared assuming the 3-parameter spectral template from \citet{draine07}.
Our model does not include emission from AGN nor from ionized gas, and for this reason we mask the \OII\ line when fitting the spectra.
Two additional free parameters describing the fraction of outlier pixels and a global scaling of the spectral uncertainty are needed to account for the imperfect data reduction (see \citealt{johnson20} for details).

\begin{deluxetable*}{lCCcCCCCC}
\tablecaption{Galaxy Properties\label{tab:sample}}
\tablewidth{0pt}
\tablehead{
 \colhead{ID} & \colhead{$z$} & \colhead{$S_{CO} \Delta v$} & Ref. & \colhead{$\log M_\mathrm{gas}/\Msun$} & \colhead{$\log \Mstar/\Msun$} & \colhead{$\mathrm{SFR}_\mathrm{UV+IR}$} & \colhead{$\mathrm{SFR}_\mathrm{spec}$} & \colhead{$\mathrm{SFR}_\mathrm{[OII]}$} \\
  & & \colhead{(Jy km/s)} & & & & \colhead{(\Msun/yr)} & \colhead{(\Msun/yr)} & \colhead{(\Msun/yr)}  
 }
\colnumbers
\startdata
EGS-20106 & 1.062 & 0.43 \pm 0.06 		& This work & 10.59 \pm 0.06 	& 11.25 & 26.9 \pm 2.0 		& 21 ^{+6} _{-5} 		& 4.7 \pm 1.8 \\
EGS-18045 & 1.012 & 0.88 \pm 0.31 		& This work & 10.70 \pm 0.15 	& 11.33 & 15.8 \pm 1.2 		& 10 ^{+5} _{-4} 		& 7.8 \pm 4.1 \\
EGS-17533 & 1.264 & 0.060 \pm 0.014 		& This work & 9.88 \pm 0.10 	& 10.78 & 14.1 \pm 4.9 		& 6.4 ^{+2.2} _{-2.0} 		& 1.6 \pm 0.6 \\
SDSS J0912+1523 & 0.747 & 1.07 \pm 0.05 & Suess+17 & 10.68 \pm 0.02 	& 11.23 & <257.0 			& 52 ^{+20} _{-20} 		& 4.6 \pm 1.4 \\
SDSS J2202-0033 & 0.657 & 0.27 \pm 0.03 & Suess+17 & 9.97 \pm 0.05 		& 11.18 & <70.8 			& 12 ^{+14} _{-9.6} 	& 2.9 \pm 1.2 \\
110509 & 0.667 & 0.24 \pm 0.04 			& Spilker+18 & 9.93 \pm 0.07 	& 11.00 & 5.8 \pm 0.8 		& 0.8 ^{+0.6} _{-0.5} 		& \nodata \\
130284 & 0.602 & 0.36 \pm 0.04 			& Spilker+18 & 10.02 \pm 0.05 	& 10.99 & 5.9 \pm 0.6 		& 2.7 ^{+0.7} _{-0.6} 		& \nodata \\
132776 & 0.750 & 0.33 \pm 0.07 			& Spilker+18 & 10.18 \pm 0.09 	& 10.98 & 6.9 \pm 1.0 		& 3.1 ^{+0.7} _{-0.6} 		& 1.6 \pm 0.2 \\
138718 & 0.656 & <0.21			 		& Spilker+18 & <9.86 			& 11.20 & 3.2 \pm 0.6 		& 2.1 ^{+0.8} _{-0.7} 		& 0.5 \pm 0.2 \\
169076 & 0.677 & <0.23 					& Spilker+18 & <9.93 			& 11.45 & 4.2 \pm 0.6 		& 0.5 ^{+0.4} _{-0.4} 		& \nodata \\
210210 & 0.654 & <0.21 					& Spilker+18 & <9.86 			& 11.38 & 2.3 \pm 0.6 		& <0.1 		& \nodata \\
211409 & 0.714 & <0.13 					& Spilker+18 & <9.73 			& 11.13 & 5.9 \pm 0.8 		& <0.1 		& \nodata \\
\enddata
\tablecomments{
(1) ID, which for our targets matches the 3D-HST identification (while the IDs used in \citealt{belli14lris} are 51081, 51106, and 22780 for, respectively, 20106, 18045, and 17533); (2) Redshift measured from rest-frame optical spectroscopy; (3) Observed line-integrated flux of CO(2-1); for EGS-18045, CO(3-2) was observed instead; (4) Reference for the CO observations; (5) Molecular gas mass; (6) Stellar mass from SED fitting; (7) SFR derived from UV+IR; (8) SFR derived from fitting photometry and spectroscopy with \texttt{Prospector} (the uncertainties span the central 68\% of the posterior distribution); (9) SFR derived from the dust-corrected \OII\ line flux (only available if the spectrum covers the rest-frame \OII\ wavelength).}
\end{deluxetable*}

The resulting star formation histories, shown in Figure~\ref{fig:fit}, confirm the diversity of the sample. EGS-20106 features a slow quenching followed by a recent rejuvenation event in which the SFR increased by more than a factor of 2, which may be related to the presence of a gas-rich satellite discussed in Section~\ref{sec:CO}; although we note that another valid interpretation is that the observed star formation history consists of the sum of a slowly quenching galaxy and a young, low-mass satellite. EGS-18045 is undergoing quenching at a slow, constant rate, but the uncertainties are large enough that we cannot rule out a rejuvenation event in this system as well.
Finally, EGS-17533 is a clear example of post-starburst galaxy with a spectrum dominated by Balmer absorption lines, for which the SFR dropped by an order of magnitude over the last $\sim500$ Myr. Consistent with the results of spectral fitting, on the $U-V$ vs $V-J$ space this object lies near (but slightly outside) the post-starburst region defined in \citet{belli19}.


\section{Results}
\label{sec:results}

\subsection{Gas Mass Fractions}

\begin{figure}[t]
\centering
\includegraphics[width=0.45\textwidth]{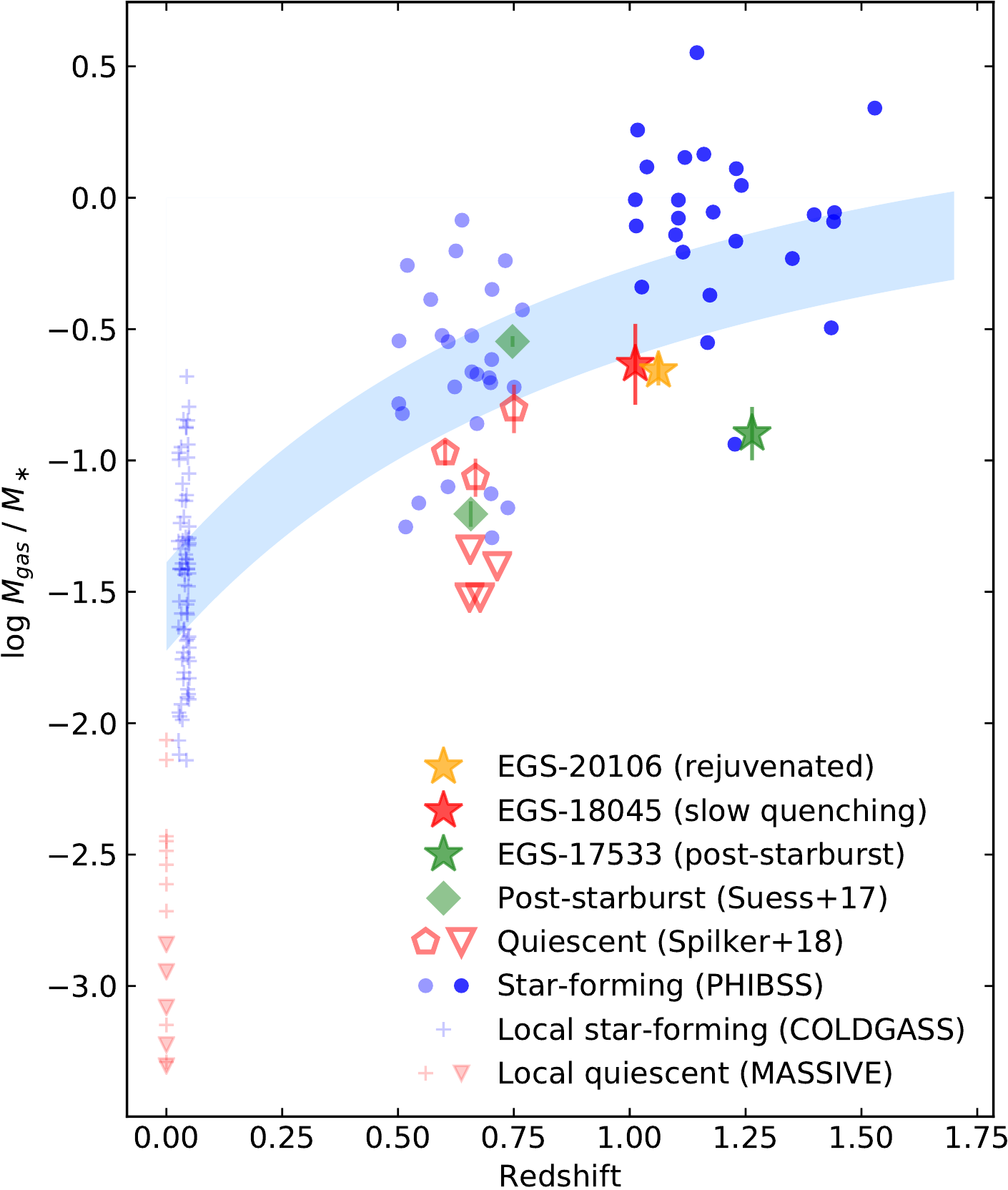}
\caption{Gas mass fraction as a function of redshift for our sample (star symbols) and other samples from the literature. Triangles represent upper limits. The shaded area corresponds to the \citet{tacconi18} scaling relation for galaxies of mass $10^{11} \Msun$ that are within 0.3 dex of the main sequence. \label{fig:gasmass}}
\end{figure}

The gas mass fraction, defined as the ratio of molecular gas mass \Mgas\ to the stellar mass \Mstar, is a fundamental property of galaxies since it relates the amount of fuel available for star formation to the integrated amount of stars formed in the past. We measure gas mass fractions of 13-23\% for our sample. To place our results in the broader context, in Figure~\ref{fig:gasmass} we show the molecular gas mass fraction as a function of redshift for galaxies in the same mass range as our sample, i.e. with $\log \Mstar/\Msun > 10.7$. At intermediate redshift, we show the sample of quiescent galaxies from \citet{spilker18} and that of post-starburst galaxies from \citet{suess17}; while for the local universe we show the MASSIVE survey \citep{davis16}. We also show, for comparison, the population of star forming galaxies at $z\sim0$ \citep[COLDGASS,][]{saintonge11a}, $z\sim0.7$ \citep[PHIBSS2,][]{freundlich19}, and $z>1$ \citep[PHIBSS,][]{tacconi13}. For a consistent comparison, we calculated the gas masses starting from the published CO fluxes and using our choice of $\alpha_\mathrm{CO}$ and $R_{J1}$.

Star-forming galaxies are substantially more gas-rich at high redshift, as illustrated by the scaling relations derived for systems on the main sequence of star formation \citep[shaded blue area,][]{tacconi18}. In the local universe, the gas mass fractions of the quiescent population are 1 to 2 orders of magnitude lower than those of star-forming systems. In the distant universe, the few gas mass fractions measured in quiescent galaxies, including ours, are higher than those measured in local quiescent systems and span a large range of values. The origin of such diversity is not clear, and may be due to selection effects and/or intrinsic physical differences among galaxies. For a meaningful comparison of the gas masses it is necessary to take the SFR of each system into account.

\subsection{Star Formation Rates and Depletion Times}

The depletion time $t_\mathrm{depl} = \Mgas / \mathrm{SFR}$ gives a measure of the efficiency of star formation and can be used to set observational constraints on models of galaxy quenching.
However, detecting low levels of SFR in high-redshift galaxies is observationally challenging, and different techniques do not always give consistent results. We attempt to alleviate this problem by using three different methods to derive the SFR; moreover, for an accurate comparison we perform the same analysis (including the \texttt{Prospector} fits) to other similar samples observed in CO(2-1): the two post-starburst systems from \citet{suess17} and seven quiescent systems from \citet{spilker18}\footnote{From the \citeauthor{spilker18} sample we exclude one galaxy (ID 74512) because its stellar absorption lines are offset by 2400 km/s from the CO and \OII\ emission lines.}. These galaxies, whose properties are listed in Table~\ref{tab:sample}, are all at $z\sim0.7$ and have publicly available photometric and spectroscopic data. 

We measure the SFR with the following methods:
\begin{enumerate}
 \item UV+IR. We estimate the rest-UV luminosity from the best-fit spectral template, and the IR luminosity from a single photometric measurement (either MIPS 24 \micron\ or WISE), adopting the average template from \citet{dale02}; the total SFR is then given by a weighted sum of the UV and IR luminosities, following \citet{bell05} and \citet{wuyts08}

 \item Spectral fit. The youngest bin in the star formation histories obtained with \texttt{Prospector} gives a measure of the SFR averaged over the last 60 Myr.
 
 \item Dust-corrected \OII. We measure the \OII\ flux by first subtracting the best-fit stellar template and then fitting a double Gaussian profile with a fixed 1:1 line ratio, since the doublet is not well resolved in the spectra. Finally, we correct for dust attenuation by using the results of the \texttt{Prospector} fit, which includes extra attenuation towards \ion{H}{2} regions (so that the attenuation for emission lines is approximately twice that for the stellar continuum). The dust-corrected fluxes can then be used to derive SFR measurements via the \citet{kewley04} calibration, converted to a Chabrier IMF.
\end{enumerate}

\begin{figure*}[ht]
\centering
\includegraphics[width=\textwidth]{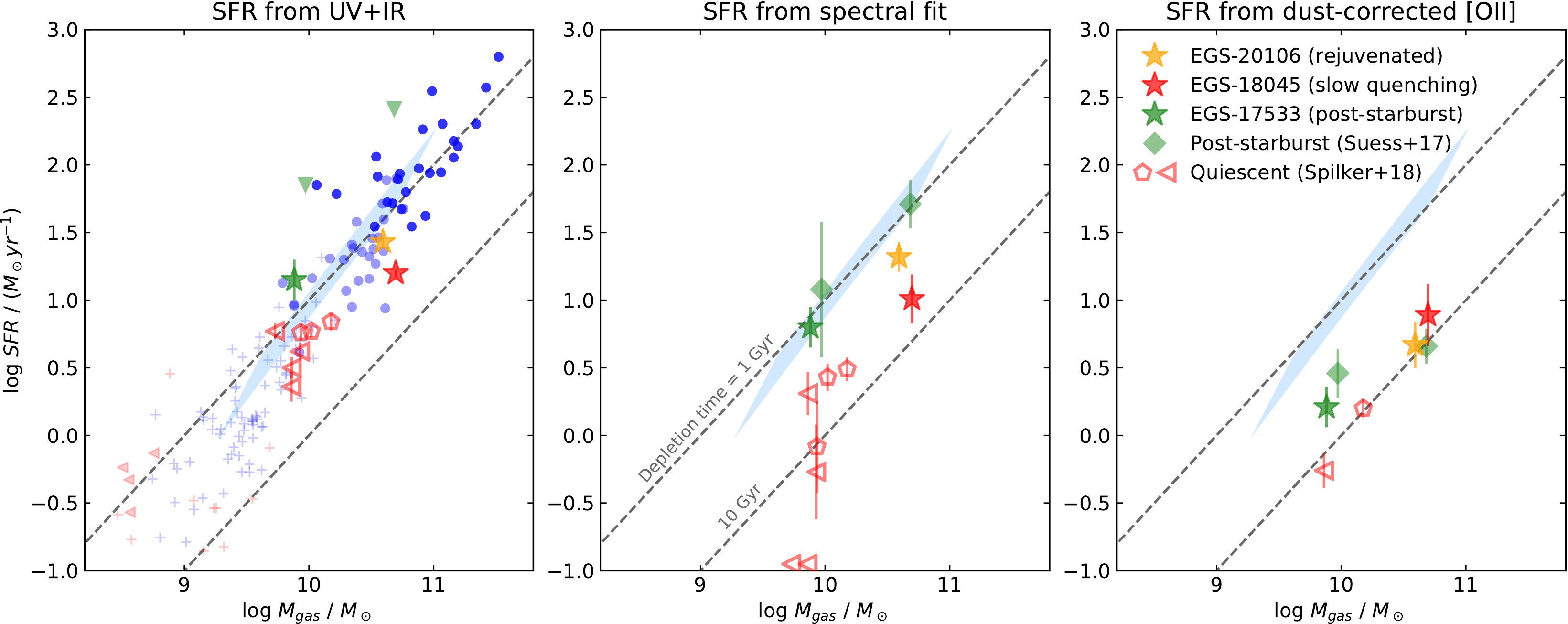}
\caption{SFR as a function of molecular gas mass, for three different methods of measuring the SFR: UV+IR (left panel), spectral fit (center), and dust-corrected \OII\ (right). Symbols as in Figure~\ref{fig:gasmass}. The dashed lines correspond to depletion times of 1 and 10 Gyr; the shaded area marks the \citet{tacconi18} scaling relations for galaxies of mass $10^{11} \Msun$ that are within 0.3 dex of the main sequence. 
\label{fig:depl}}
\end{figure*}

In Figure~\ref{fig:depl} we show the relation between the molecular gas mass and the SFR obtained with the three methods. For quiescent galaxies there are clear systematic offsets between the different techniques: the UV+IR SFRs are about two times larger than the spectral fit values, and 6 times larger than the dust-corrected \OII\ measurements. The diagonal dashed lines mark constant values of depletion time, showing that the SFR discrepancy has a strong impact on the depletion time measurements. The UV+IR method gives similar depletion times for quiescent and star-forming galaxies, while the dust-corrected \OII\ SFRs yield much longer depletion times for quiescent systems, of the order of 10 Gyr.

Recent results have shown that the UV+IR method consistently overestimates the SFR in high-redshift quiescent galaxies because of the contribution of relatively older stars to dust heating \citep[e.g.,][]{hayward14,leja19sed}. This is confirmed by our \texttt{Prospector} fits, according to which about half of the observed IR flux is due to stars older than 1 Gyr.
We therefore deem the spectral fit results more robust than the UV+IR values.
On the other hand, the SFR from dust-corrected \OII\ is consistently smaller than that from the spectral fitting, likely because of an inadequate dust correction -- this may be due to the inability to capture the \emph{nebular} dust attenuation from fitting the \emph{stellar} emission. Moreover, both the \OII\ line and the IR flux can be contaminated by AGN emission; however none of the galaxies considered here show signs of AGN activity, according to their \emph{Spitzer} or \emph{WISE} colors \citep{donley12, stern12}.
We conclude that the least biased SFR measurements come from spectral fitting. 
We note that these SFRs are substantially different from those initially used to select our sample, which were obtained via SED fitting with exponentially declining star formation histories. The large degree of flexibility built into our \texttt{Prospector} model should guarantee results that are much more robust than those of standard SED fitting. However, we caution that a small number of assumptions are still involved in the fits; for example, the size of the youngest age bin used for measuring the SFR is somewhat arbitrary. We verified that by using 30 Myr instead of 60 Myr for the youngest age bin our results are nearly unchanged (i.e., the median absolute change of the SFR is 0.05 dex). Moreover, our model implicitly assumes a moderate degree of smoothness in the star formation history; using models that allow short bursts may yield substantially different SFR measurements for post-starburst galaxies (\citealt{french18, wild20}; K. Suess et al. 2021, in preparation).

Adopting the results of spectral fitting we obtain a wide range of depletion times for quiescent galaxies, with values spanning from $\sim1$ to $\sim10$ Gyr. Interestingly, the three post-starburst galaxies, shown in green in Figure~\ref{fig:depl}, appear to have depletion times that are short and fully consistent with the star-forming main sequence, while older and more quiescent systems have longer depletion times.

\section{Discussion}

Our results confirm the surprising diversity of molecular gas content in quiescent galaxies at high redshift \citep[e.g.,][]{spilker18,bezanson19}, and suggest that differences in the depletion time may trace different stages of galaxy quenching.
Our observations are consistent with a simple picture in which rapid quenching is caused by the sudden heating or removal of molecular gas, and is not associated to a change in the efficiency of star formation. Alternatively, the cold gas could have been exhausted in a rapid burst: in this case the depletion time would be \emph{shorter} than for typical star-forming galaxies but only for a brief period immediately preceding the post-starburst phase. Another intriguing possibility is that the post-starburst phase represents a stochastic fluctuation in the life of main-sequence galaxies, which may or may not be followed by a complete quenching \citep{tacchella20}. 

An alternative class of scenarios involves the suppression of the star formation efficiency, due for example to the stabilizing effect of a massive stellar bulge \citep{martig09}. Our results rule out this mechanism as a cause for rapid quenching; however the long depletion times observed in older quiescent systems may suggest that these processes become increasingly relevant as galaxies age \citep[consistent with the analysis of dust emission by][]{gobat18}. Since the growth of massive bulges starts when galaxies are still star-forming \citep{genzel14quenching}, this may represent a slow quenching route, alternative to the quenching channel responsible for the formation of post-starburst systems \citep{belli19}. 

Ultimately, larger samples are needed to confirm a possible trend between depletion time and quenching stage. Sample size is, however, not the only limitation of current studies. First, the conversion of CO fluxes to molecular gas masses is still highly uncertain for quiescent galaxies. Secondly, our study highlighted the critical role played by the method adopted to measure the SFR. As deeper millimeter observations are able trace the molecular gas content in galaxies at increasingly high redshift and low SFR, this will become a fundamental issue. Finally, the origin of the gas reservoir may be different for different quiescent galaxies, further complicating the interpretation; in our small sample we have evidence for recent accretion due to a minor merger in one galaxy, and a common origin for the gas and stellar content in another \citep[see also][]{spilker18, hunt18}.
Only by performing a detailed study of the stellar and gas content on a large sample of quiescent galaxies will we finally be able to understand the role played by the cold gas reservoir in galaxy quenching. \\

We thank the referee for an insightful report, and Wren Suess for many helpful discussions.
S.B. acknowledges support from the Clay Fellowship. This work is based on observations carried out under projects number W18DF and W19CJ with the IRAM NOEMA Interferometer. IRAM is supported by INSU/CNRS (France), MPG (Germany) and IGN (Spain). The spectral fits were carried out on \texttt{Hydra}, the Smithsonian Institution High Performance Cluster (SI/HPC).


\bibliography{sirio}{}
\bibliographystyle{aasjournal}

\end{document}